\documentclass[preprint,showpacs,preprintnumbers,amsmath,amssymb]{revtex4}

\usepackage{graphicx}
\usepackage{dcolumn}
\usepackage{bm}

\begin{document}


\title{Phase-shift in vapor cell and compact cold-atom frequency standards}

\author{Aldo Godone and Salvatore Micalizio}
\affiliation{%
INRIM, Istituto Nazionale di Ricerca Metrologica, Strada delle Cacce 91, 10135, Torino, Italy\\
}

\date{\today}

\begin{abstract}
We report on a theoretical analysis of the phase-shift in compact atomic clocks working either with cold or thermal atoms. It is well known that in a microwave cavity with electromagnetic losses, a traveling wave adds to the standing wave of a given resonant mode. We calculate the spatial varying phase related to this travelling wave for two geometries of interest in clock applications, the cylindrical cavity and the spherical cavity. Due to their motion, the atoms probe different regions of the cavity and then experience different phases of the interrogating microwave field. We show that this combination of atomic motion and spatial changing phase results in a phase-shift of the clock frequency, well known in primary frequency standards, that can affect the metrological performances also of vapor cell and compact cold-atom clocks. In the latter case, we evaluate the phase-shift for a space clock and a ground clock.
\end{abstract}

\maketitle

\section{Introduction}
In primary atomic frequency standards, the spatial phase variations of the microwave field exciting the atoms cause a shift of the observed clock transition from its unperturbed value. This phase shift has been extensively studied for a long time and in different atomic clock setups, including Cs \cite{bauch} and Mg \cite{bava} thermal beams and, more recently, atomic fountains \cite{eftf2006, weyers}. It represents an important source of systematic error in the evaluation of primary frequency standards accuracy budget \cite{li, eftf2010, szymaniec, levi, heavner, ashby}.

\
From a physical point of view, the phase shift is strictly related to the losses in the conducting walls of the resonant microwave cavity sustaining the (interacting) electromagnetic field and in the dielectric medium that may be contained in the cavity itself. In the limit of small losses well satisfied in high resolution spectroscopy arrangements, this field can be considered as a superposition of a standing wave, which results from the cavity geometry, and a small amplitude travelling wave whose intensity is proportional to the losses \cite{vecchi, khursheed}. The field has then a spatially varying phase and since the atoms probe different regions of the cavity these phase variations give rise to a distributed cavity phase shift of the clock frequency \cite{li2004, li2010}.

\
In vapor cell frequency standards, the phase shift has not been considered because in those clocks accuracy is not an important issue and its possible impact on the frequency stability has been assumed marginal; in fact, other effects are believed to limit the clock stability, such as the light shift or the thermal sensitivity related to the atom-buffer gas collisions \cite{godone2015}. However, the recently developed pulsed optically pumped (POP) Rb cell clock reached a medium-term stability in the $10^{-15}$ range \cite{micalizio2012} and a similar result was achieved by a lamp-pumped Rb clock for space applications \cite{mei}; it is then reasonable to wonder whether the phase shift may play a role in affecting the clocks' stability at that level. In addition, there is an increasing interest for a new class of cold-atom clocks that aim at joining good accuracy and stability performances with features that are typical of cell clocks, like compactness and reliability. In these relatively small devices a single cavity is used for all the clock's operation stages, including atom cooling and interrogation. The atom cooling is performed either by applying a magneto-optical trap (MOT) \cite{muller2011} or by using isotropic light \cite{esnault2010, desarlo2014, pengliu, huadong}.  Also for these compact cold atom clocks it is necessary to evaluate the possible impact of the phase shift on both the accuracy, which is expected at the level of $10^{-14}$ or better, and the stability. In this paper we theoretically study the phase shift in the POP cell clock and in compact cold-atom clocks and, although so far disregarded, we demonstrate that its effect can play an important role in limiting their metrological performances.

\
In this work, we will study in detail the cylindrical cavity working on the TE$_{011}$ mode (section II) and the spherical cavity operating on the TE$_{110}$ (section III). The first cavity is widely adopted in vapor cell frequency standards with buffer gas and also is some cold-atom clocks, whereas the spherical cavity is used in the compact cold-atom clocks described in \cite{esnault2010} and in \cite{desarlo2014}. We notice that in these compact cold-atom clocks no buffer gas is used and therefore their metrological properties, including the possibility of providing a significant accuracy, are typical of cold-atom frequency standards.

\
To evaluate the phase shift we will exploit the unperturbed analytical expressions of the electromagnetic modes as they result from the solution of the Maxwell's equations with the boundary conditions of a conventional cylindrical or spherical cavity; this means that we will neglect the effect of the cell (when present) to define the shape of the modes sustained by the cavity, as well as the possible presence of holes and feeds. A quantitatively more precise evaluation clearly requires a numerical integration of Maxwell's equations for the specific geometry under consideration. However, the level of approximation here adopted is sufficient to give a description and the order of magnitude of the phase shift and to evaluate its impact on the clocks considered in this paper (Section IV).

\
Conclusions are reported in Section V.
\section{Cylindrical cavity}

In Fig.1 we show the cylindrical configuration we consider in this section.
\begin{figure}[!]
\begin{center}
\includegraphics[height=200pt]{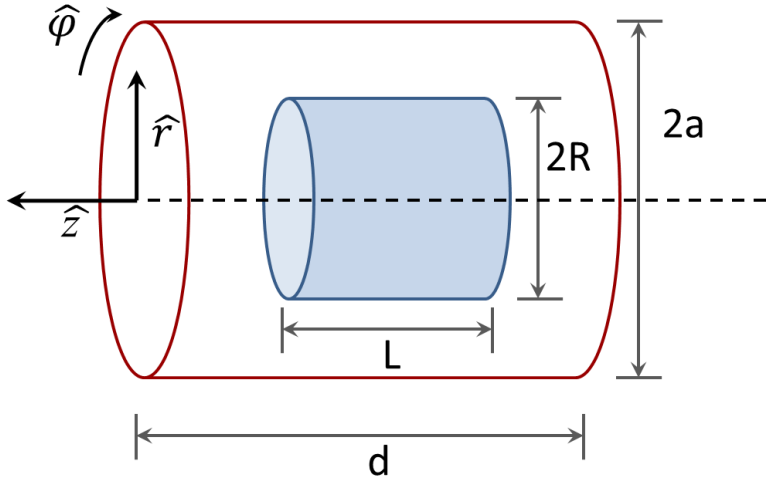}
\caption{(Color online) Cylindrical cavity; $\hat{r}$, $\hat{\varphi}$ and $\hat{z}$ are the versors identifying the cylindrical coordinate system. $2 a$ and $d$ are the diameter and the length of the cavity respectively. Similarly, $2 R$ and $L$ are the diameter and the length of the cell.}\label{cyl_cavity}
\end{center}
\end{figure}
\
The atomic vapor is confined in a region of space inside the cavity either by a dielectric cell or by a laser cooling technique. The vapor and the cell (if present) are supposed not to alter significantly the electromagnetic mode inside the cavity. We assume the cavity sustains a microwave field resonating at the atomic frequency and operating at the TE$_{011}$ mode. This mode is commonly adopted in many experimental implementations of atomic clocks since the cavity axis can conveniently identify a quantization axis. According to selection rules, the so called $m_{F}=0-m_{F}=0$ clock transition can be excited by a microwave magnetic field if its field lines are parallel to the quantization axis: the TE$_{011}$ mode satisfies this requirement and in addition guarantees a high (intrinsic) cavity quality factor. In the POP clock, the wave-vector of the laser used to pump and detect the atoms is parallel to the cavity's axis, so that the cylindrical symmetry is kept.

\
In this section we are concerned with the traveling wave phase that, due to electromagnetic losses, adds to the standing wave sustained by the cavity. It is convenient to start with the cylindrical waveguide operating in the TE$_{01}$ mode and whose side view is shown in Fig. 2.

\begin{figure}[!]
\begin{center}
\includegraphics[height=200pt]{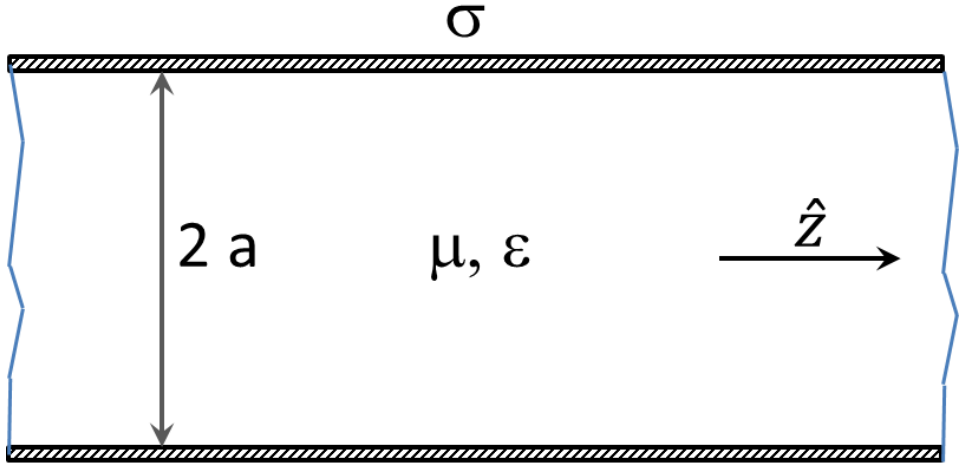}
\caption{(Color online) Cylindrical waveguide of diameter equal to $2a$. $\mu$ and $\epsilon$ are the magnetic permeability and dielectric constant of the inner medium respectively; $\sigma$ is the conductivity of the metallic walls.}\label{cyl_waveguidde}
\end{center}
\end{figure}
\
According to \cite{Marcuvitz}, the TE$_{01}$ field components can be written as ($j=\sqrt{-1}$):


\begin{subequations}
\label{modes_te10}
\begin{eqnarray}
E_{\varphi}&=&-\frac{1}{\sqrt{\pi}} \frac{V(z)}{a \ J_{0}(x_{01}')} J_{1}\left(\frac{x_{01}' r}{a}\right)\label{equationa}
\\
H_{z}&=& -j \frac{1}{\sqrt{\pi}} \frac{\lambda \ x_{01}'}{2\pi a} \frac{1}{Z_{0}} \frac{V(z)}{a \ J_{0}(x_{01}')} J_{0}\left(\frac{x_{01}' r}{a}\right)\label{equationb}
\\
H_{r}&=& \frac{1}{\sqrt{\pi}} \frac{I(z)}{a J_{0}(x_{01}')} J_{1}\left(\frac{x_{01}' r}{a}\right) \label{equationc}
\end{eqnarray}
\end{subequations}

where $J_{\nu}(x)$ is the Bessel function of the first type and order $\nu$, $x_{01}'= 3.832$ ($J_{0}'(x_{01}')=0$), $a$ is the cavity radius, $\lambda$ is the free space microwave wavelength and $Z_{0}= \sqrt{\mu/\epsilon}$ is the characteristic impedance of the medium. In previous equation, $V(z)$ and $I(z)$ are the mode voltage and the mode current respectively; their expressions are the key point for the evaluation of the phase of the field we are interested to in this paper.

\
The cylindrical waveguide is a uniform transmission line where the longitudinal eigenfunctions along the propagation direction ($z$-axis) are \cite{Marcuvitz}:

\begin{subequations}
\label{v_modes} 
\begin{eqnarray}
^{+}V(z)&=&^{+}V_{0} \ e^{-\gamma (z-z_{0})}\label{equationa}
\\
^{-}V(z)&=&^{-}V_{0} \ e^{+\gamma (z-z_{0})}\label{equationb}
\end{eqnarray}
\end{subequations}

where $^{+}V_{0}$ ($^{-}V_{0}$) is the incident (reflected) complex voltage amplitude in $z=z_{0}$ and $\gamma$ is the (complex) propagation constant:

\begin{equation}
\gamma=\alpha+j  k_{z}
\end{equation}

$\alpha$ being the attenuation coefficient related to the losses in the metallic surface and in the dielectric inner medium and $k_{z}$ the wave number in the waveguide.

\
We remind that:

\begin{equation}
k_{z}=\frac{2 \pi}{\lambda_{g}}=\sqrt{k^{2}-k_{c}^2}
\end{equation}

where $\lambda_{g}=\frac{\lambda}{\sqrt{1-(\lambda/\lambda_{c})^2}}$ is the guided wave number, $k=2\pi/\lambda$ is the free space wavenumber in the medium inside the cavity, $k_{c}=2\pi/\lambda_{c}$ is the cutoff wavenumber and $\lambda_{c}$ the corresponding wavelength that for the TE$_{01}$ is given by $\lambda_{c}=2\pi a /x_{01}'$.

\
The attenuation coefficient can be expressed with a good approximation as the sum of two contributions:

\begin{equation}
\alpha=\alpha_{w}+\alpha_{i}
\end{equation}
where $\alpha_{w}$ is the attenuation coefficient due to losses in the metallic cylindrical surface and is given by \cite{collin}:

\begin{equation}
\alpha_{w}=\frac{R_{S}}{a Z_{0}}\left(\frac{\lambda}{\lambda_{c}}\right)^2\frac{1}{\sqrt{1-(\lambda/\lambda_{c})^2}}
\end{equation}

where $R_{S}=\sqrt{\frac{\omega \mu}{2 \sigma}}$ is the surface resistance and $\omega$ the microwave (angular) frequency.

\
The attenuation coefficient $\alpha_{i}$ is related to the losses in the inner medium and is given by \cite{Marcuvitz}:
\begin{equation}
\alpha_{i}=\frac{\pi\lambda_{g}}{\lambda^{2}} \Im\epsilon_{r}
\end{equation}

where $\Im$ stands for imaginary part and $\epsilon_{r}=\epsilon/\epsilon_{0}$. The previous equation is valid in the limit of low electric losses ($\Im\epsilon_{r}\ll 1$) and with no magnetic losses ($\Im\mu=0$).

\
The scattering description commonly developed in the transmission line theory turns out particularly effective to the problem at hand; in particular the mode voltage at point $z$ can be written as:

\begin{equation}\label{mode_voltage}
V(z)=  \mathop{^{+}V(z)}+ ^{-}V(z)=\mathop{^{+}V(z)} \left[1+ \mathop{^{v}\Gamma(z)}\right]
\end{equation}

where the voltage reflection coefficient is defined as:

\begin{equation}\label{reflection_coeff}
^{v}\Gamma(z)=\frac{^{-}V(z)}{^{+}V(z)}=\mathop{^{v}\Gamma(z_{0})} \ e^{2\gamma(z-z_{0})}
\end{equation}

Taking into account Eqs. (\ref{v_modes}) and definition (\ref{reflection_coeff}), Eq. (\ref{mode_voltage}) yields for the mode voltage:
\begin{equation}\label{mode_voltage_2}
V(z)=\mathop{^{+}V_{0}}\left[e^{-\gamma(z-z_{0})}+\mathop{^{v}\Gamma(z_{0})} \ e^{+\gamma(z-z_{0})}\right]
\end{equation}

This relation describes a transmission line terminated in $z=z_{0}$ by a loaded impedance $Z_{L}$, as reported in Fig. 3.

\begin{figure}[!]
\begin{center}
\includegraphics[height=200pt]{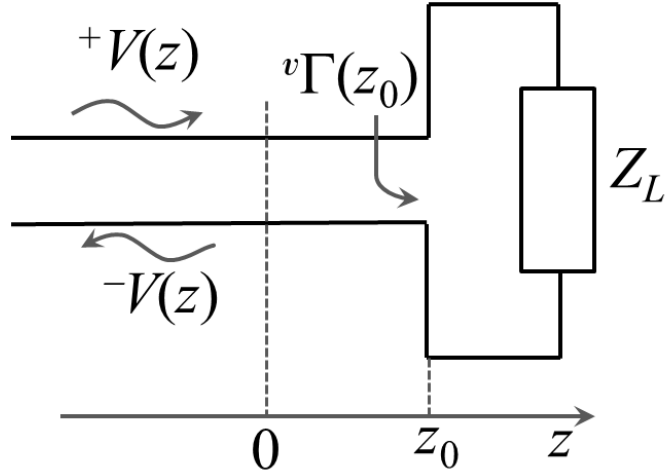}
\caption{Equivalent transmission line circuit corresponding to Eq. (\ref{mode_voltage_2}) of the text.}\label{trasm_line}
\end{center}
\end{figure}
The corresponding reflection coefficient $^{v}\Gamma(z_{0})$ is given by \cite{Marcuvitz, collin}:

\begin{equation}\label{reflection_coeff2}
^{v}\Gamma(z_{0})=\frac{z_{L}-1}{z_{L}+1}
\end{equation}

where $z_{L}$ is the loaded impedance normalized to the mode transverse impedance $Z_{TE}$ ($Z_{TE}=Z_{0}\lambda_{g}/\lambda$):

\begin{equation}
z_{L}=\frac{Z_{L}}{Z_{TE}}=\frac{Z_{L}}
{Z_{0}}\frac{\lambda}{\lambda_{g}}
\end{equation}

Finally, inserting Eq. (\ref{reflection_coeff2}) into Eq. (\ref{mode_voltage_2}) we obtain the expression of the mode voltage ($V_{0}\equiv ^{+}V(z=0)$):
\begin{equation}\label{mode_voltage_3}
V(z)=\frac{2V_{0}e^{-\gamma z_{0}}}{z_{L}+1}\left\{z_{L}\cosh[\gamma (z-z_{0})]-\sinh[\gamma (z-z_{0})]\right\}
\end{equation}

This equation is strictly analogous to that reported in \cite{Vanier} for a rectangular waveguide.
The result expressed by Eq. (\ref{mode_voltage_3}) for a terminated waveguide can be very conveniently used to define a resonant microwave cavity as follows: i) $z_{0}$ is set equal to $d/2$ ($z_{0}=d/2$) so that the coordinate system is centered in the cavity center; ii) $Z_{L}$ is set equal to $Z_{S}$, where $Z_{S}$ is the surface impedance of the cavity end-cap (placed in $z=z_{0}=d/2$); for metal conductors $Z_{S}=R_{S}(1+j)$ (Leontovich boundary conditions); iii) a boundary condition for the field $H_{z}$ is specified at $z=-d/2$. In particular, in the limit of low losses we have $\alpha |z-z_{0}|\ll 1$ and $r_{S}=\frac{R_{S}}{Z_{TE}}\ll 1$ and Eq. (\ref{mode_voltage_3}) can be simplified in the form:

\begin{equation}\label{mode_voltage_4}
V(z) \approx 2V_{0}\sin k_{z}(z-z_{0}) \ e^{i \phi(z)}
\end{equation}

where the phase $\phi(z)$ is specified in the following. If we suppose that the source of the field is in $z= -d/2$ we have $|H_{z}(z=-d/2)|=0$ which implies $|V(z=-d/2)|=0$ and then $k_{z}=\pi/d$.
With this condition, Eq. (\ref{mode_voltage_3}) yields:

\begin{equation}\label{mode_voltage_5}
V(z) = 2V_{0}\cos \frac{\pi}{d}z \ e^{i \phi(z)}
\end{equation}

The phase factor $\phi(z)$ is given by (neglecting additive constants independent from position):

\begin{equation}\label{phase}
\tan\phi(z)= \frac{\left[r_{S}-\alpha(z-d/2)\right] \tan \frac{\pi}{d}z}{1+r_{S}\tan\frac{\pi}{d}z}
\end{equation}

Equations (\ref{mode_voltage_5}) and (\ref{phase}) describe the steady state wave of the resonant mode corrected by the phase factor $e^{i \phi}$ which accounts for losses. It is worth noting that in Eq. (\ref{phase}) $r_{S}$ refers to the cavity end-cap, whereas $\alpha$ refers to the waveguide and in principle they could be composed of different materials.

\
The mode current $I(z)$ is easily derived from the transmission line equation \cite{Marcuvitz, collin}:

\begin{equation}
\frac{d V(z)}{dz}=-\gamma Z_{TE} \ I(z)
\end{equation}
From Eq. (\ref{mode_voltage_5}) it turns out:

\begin{equation}\label{mode_current_cyl}
I(z)=-j \frac{1}{Z_{0}}2 V_{0}\frac{\lambda_{0}}{2 d}\sin\frac{\pi}{d}z \ e^{j[\phi(z)+\psi(z)]}
\end{equation}

where we have taken into account that at resonance ($\lambda= \lambda_{0}$) the impedance of the mode $Z_{TE}$ is $Z_{TE}=Z_{0}\frac{2d}{\lambda_{0}}$.

\
The phase term $\psi(z)$ is given by:

\begin{equation}\label{psi}
\psi(z)=\arg\left\{\frac{\frac{d\phi}{dz}\cot\frac{\pi}{d}z-j\frac{\pi}{d}}{\alpha+j\frac{\pi}{d}}\right\}
\end{equation}

Inserting Eqs. (\ref{mode_voltage_5}) and (\ref{mode_current_cyl}) into (\ref{modes_te10}) and normalizing to $A_{H}$ the value of $H_{z}$ in the cavity center, we obtain the expression of the field components for the TE$_{011}$ mode:
\begin{subequations}
\label{modes_te011_def}
\begin{eqnarray}
E_{\varphi}&=&-j A_{H} Z_{0} \frac{2 \pi}{\lambda_{0}}\frac{a}{x_{01}'} J_{1}\left(\frac{x_{01}' r}{a}\right) \cos\frac{\pi}{d}z \ e^{j \phi(z)} \label{equationa}
\\
H_{z}&=& A_{H} J_{0}\left(\frac{x_{01}' r}{a}\right) \cos\frac{\pi}{d}z \ e^{j \phi(z)} \label{equationb}
\\
H_{r}&=& A_{H} \frac{a}{x_{01}'} \frac{\pi}{d}  J_{1}\left(\frac{x_{01}' r}{a}\right)\sin\frac{\pi}{d}z \ e^{j[\phi(z)+\psi(z)]}  \label{equationc}
\end{eqnarray}
\end{subequations}

Unless the phase factor $\phi(z)$, Eqs. (\ref{modes_te011_def}) reproduce the expressions already reported in \cite{godone2011}.
The behavior of $\phi(z)$ is reported in Fig. 4 for two values of $r_{S}$; the figure refers to a cylindrical cavity resonating at $\nu_{0}=7.7$ GHz filled with a medium with no dielectric losses ($\alpha_{i}=0$) and with diameter $2a=52$ mm and length $d=49$ mm.

\begin{figure}[!]
\begin{center}
\includegraphics[height=250pt]{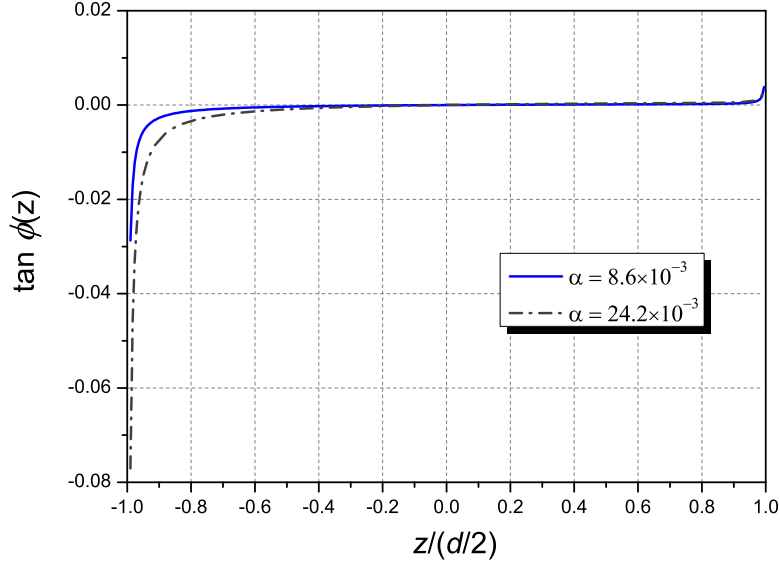}
\caption{(Color online) Plot of $\tan\phi(z)$ versus $z$ for two values of $\alpha$. continuous line: $\alpha=8.6 \times 10^{-3} \ \textrm{m}^{-1}$ corresponding to a cylindrical waveguide in Mo; dash-dot line: $\alpha=24.2\times10^{-3} \ \textrm{m}^{-1}$ corresponding to a waveguide in Ti. In both cases, the waveguide is closed by two end caps in Al ($r_{S}=3.1 \times 10^{-5}$).}\label{tanphi_cyl_cavity}
\end{center}
\end{figure}

In the central region of the cavity, the phase is almost linear and Eq. (\ref{phase}) can be more conveniently written as:

\begin{equation}\label{phase_2}
\phi(z)\approx \left[r_{S}-\alpha(z-d/2)\right] \tan \frac{\pi}{d}z
\end{equation}

as similarly reported in \cite{Vanier} for a rectangular cavity.

We do not discuss here the phase term $\psi(z)$ characterising the current mode in Eq. (\ref{mode_current_cyl}), since this phase term does not affect $H_{z}$ that, as previously discussed, is the only field component involved in the clock operation.

\section{Spherical cavity}

In this section we consider the spherical symmetry configuration reported in Fig. 5. The atoms are supposed to be confined in a spherical region of radius $R$ around the center of the cavity. Also in this case, we assume the electromagnetic mode is not significantly affected by the presence of the atomic vapor. In addition, the cavity is filled with a medium with no losses, so that $\Im(\epsilon_{r})=0$ and $\Im(\mu)=0$. In practical applications, the medium is simply the vacuum.
\begin{figure}[!]
\begin{center}
\includegraphics[height=200pt]{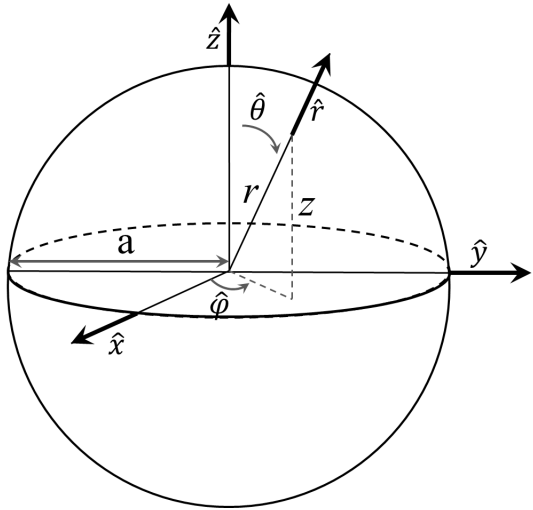}
\caption{Spherical cavity of radius $a$; $\hat{r}$, $\hat{\theta}$ and $\hat{\varphi}$ are the unit vectors of the spherical coordinate system; $\hat{x}$, $\hat{y}$ and $\hat{z}$ are the unit vectors of the Cartesian coordinate system.}\label{spher_cav}
\end{center}
\end{figure}

\
The electromagnetic mode we consider is the TE$_{110}$ commonly adopted in frequency standards based on isotropic laser cooling; this mode guarantees a highly uniform microwave field in the central part of the cavity.

\
Similarly to the cylindrical configuration analyzed in previous section, we initially consider the spherical waveguide (spherical mode TE$_{10}$): the free space is regarded as transmission line where the propagation direction is along the radius $r$. For a fixed $r$, transverse cross sections are spherical surfaces described by the coordinates $\theta$ and $\phi$. The spherical cavity is then conveniently regarded as a spherical wave guide terminated by a spherical end-cap with electrical conductivity $\sigma$.

\
As reported by \cite{Marcuvitz}, the field component of the spherical TE$_{10}$ mode are:

\begin{subequations}
\label{modes_te10_spher}
\begin{eqnarray}
E_{\varphi}&=&-\sqrt{\frac{3}{8\pi}} \frac{V(r)}{r} \sin\theta\label{equationa}
\\
H_{\theta}&=& \sqrt{\frac{3}{8\pi}} \frac{I(r)}{r} \sin\theta \label{equationb}
\\
H_{r}&=& -j \frac{1}{Z_{0}}\sqrt{\frac{3}{8\pi}} \frac{V(r)}{r} \frac{2}{k \ r} \cos\theta\label{equationc}
\end{eqnarray}
\end{subequations}

where $V(r)$ and $I(r)$ are the mode voltage and current respectively, $Z_{0}$ is the characteristic impedance of the medium and $k$ is free-space wave number.

\
The spherical waveguide is a non-uniform transmission line whose wavefunctions for the TE$_{10}$ mode in the propagation direction identified by the versor $\hat{\textbf{r}}$ are \cite{Marcuvitz, stratton}:

\begin{subequations}
\begin{eqnarray}
^{+}V(r)=\mathop{^{+}V_{0} } \hat{H}_{1}^{(1)}(kr_{0})\hat{H}_{1}^{(2)}(kr)\label{equationa}
\\
^{-}V(r)=\mathop{^{-}V_{0}} \hat{H}_{1}^{(2)}(kr_{0})\hat{H}_{1}^{(1)}(kr)\label{equationb}
\end{eqnarray}
\end{subequations}

where $^{+}V(r)\left[1+\frac{1}{(k r_{0}^2)}\right]$ and $^{-}V(r)\left[1+\frac{1}{(k r_{0}^2)}\right]$ are the outgoing and ingoing complex mode-voltage amplitudes at $r=r_{0}$, respectively; $\hat{H}_{\nu}^{(1)}(x)$ and $\hat{H}_{\nu}^{(2)}(x)$ are spherical Hankel functions of order $\nu$.

\
In the definition given in \cite{Marcuvitz}, these functions are expressed as:

\begin{subequations}
\begin{eqnarray}
\hat{H}_{\nu}^{(2)}=\hat{J}_{\nu}(x)-j \hat{N}_{\nu}(x)=-\hat{h}_{\nu}(x)e^{-j \hat{\eta}_{\nu}(x)}\label{equationa}
\\
\hat{H}_{\nu}^{(1)}=\hat{J}_{\nu}(x)+j \hat{N}_{\nu}(x)=-\hat{h}_{\nu}(x)e^{+j \hat{\eta}_{\nu}(x)}\label{equationb}
\end{eqnarray}
\end{subequations}

where $\hat{J}_{\nu}(x)$ and $\hat{N}_{\nu}(x)$ are related to the spherical Bessel functions of the first type ($j_{\nu}(x)$) and of the second type ($y_{\nu}(x)$), respectively \cite{abramovitz}:

\begin{subequations}
\begin{eqnarray}
\hat{J}_{\nu}(x)=x \ j_{\nu}(x)\label{equationa}
\\
\hat{N}_{\nu}= x \ y_{\nu}(x)\label{equationb}
\end{eqnarray}
\end{subequations}
\
The amplitude $\hat{h}_{\nu}(x)$ and the phase $\hat{\eta}_{\nu}(x)$ of the spherical Hankel functions emphasize the analogy with the exponential functions typical of the uniform transmission lines and are given by:

\begin{subequations}
\begin{eqnarray}
\hat{h}_{\nu}(x)= \sqrt{\hat{J}_{\nu}^{2}(x)+\hat{N}_{\nu}^{2}(x)}
\\
\hat{\eta}_{\nu}(x)= (\nu+1)\frac{\pi}{2}+\arctan\frac{\hat{N}_{\nu}(x)}{\hat{J}_{\nu}(x)}
\end{eqnarray}
\end{subequations}

A scattering description can be developed also in this case in exact analogy to the uniform lines. Specifically, the fundamental relation linking the reflection coefficients of two arbitrary points $r$ and $r_{0}$ of the non-uniform transmission line is:

\begin{equation}
^{v}\Gamma(r)=\mathop{^{v}\Gamma(r_{0})} \ e^{2j\left[\hat{\eta}_{1}(kr)-\hat{\eta}_{1}(kr_{0})\right]}
\end{equation}

The mode voltage turns out:


\begin{equation}\label{voltage_spher}
\begin{split}
V(r)=&^{+}V(r) \left[1+^{v}\Gamma(r)\right]\\
=&^{+}V_{0} \ \hat{h}_{1}(kr)\hat{h}_{1}(kr_{0})\left\{e^{-j [\hat{\eta}_{1}(kr)-\hat{\eta}_{1}(kr_{0})]}+\mathop{^{v}\Gamma(r_{0})} e^{+j\left[\hat{\eta}_{1}(kr)-\hat{\eta}_{1}(kr_{0})\right]}\right\}
\end{split}
\end{equation}

The relation between the load impedance and the reflection coefficient at $r=r_{0}$ is similar to Eq. (\ref{reflection_coeff2}) but now the normalization is done with respect to the characteristic impedance of free space $Z_{0}$. Then we can write:


\begin{equation}\label{mode_voltage_spher0}
V(r)=\frac{2 \ ^{+}V_{0}}{z_{L}+1} \hat{h}_{1}(kr)\hat{h}_{1}(kr_{0})\left\{z_{L}\cos\left[\hat{\eta}_{1}(kr)-\hat{\eta}_{1}(kr_{0})\right]-j \sin\left[\hat{\eta}_{1}(kr)-\hat{\eta}_{1}(kr_{0})\right]\right\}
\end{equation}

Similarly to what has been done in the previous section (see Eq. \ref{mode_voltage_5}), we aim at expressing the mode voltage as a standing wave times a phase factor which accounts for (low) losses. For a spherical cavity, it is more convenient to consider the following two steps.

\
Firstly, the spherical cavity without losses is realized by simply terminating the spherical waveguide at $r_{0}=a$ with an ideal short-circuit, that is  $^{v}\Gamma(r_{0})=-1$. The mode voltage $V(r)$ given by Eq. (\ref{voltage_spher}) becomes:

\begin{equation}
V(r)=2j \ ^{+}V_{0}\left\{\hat{J}_{1}(kr)\hat{N}_{1}(ka)-\hat{N}_{1}(kr)\hat{J}_{1}(ka)\right\}
\end{equation}

In addition, we impose that the mode voltage is finite everywhere, $V(r)\neq \infty$ for each $r$. Because of the divergence of $\hat{N}_{1}(kr)$ in $r=0$, this condition implies $\hat{J}_{1}(ka)=0$, that is $k=\xi_{11}/a$, where $\xi_{11}=4.4934$ is the first zero (not null) of $j_{1}(x)$, and then also of $\hat{J}_{1}(x)$.

\
The TE$_{110}$ standing wave in terms of the voltage mode can then be written as:

\begin{equation}\label{mode_voltage_spher}
V(r)=2 j \mathop{^{+}V_{0}}\hat{N}_{1}(\xi_{11})\hat{J}_{1}(\xi_{11}r/a)
\end{equation}

The second step concerns the phase associated to losses. Specifically, we assume for the short-circuit a normalized surface impedance given by $z_{L}=z_{S}=(1+j)R_{S}/Z_{0}$ and taking into account Eqs. (\ref{mode_voltage_spher0}) and (\ref{mode_voltage_spher}) we find out:

\begin{equation}\label{mode_voltage_spher2}
V(r)=2 j ^{+}V_{0}\hat{N}_{1}(\xi_{11})\hat{J}_{1}(\xi_{11}r/a)=V_{0}\hat{J}_{1}(\xi_{11}r/a)e^{j\phi(\xi_{11}r/a)}
\end{equation}

where the phase factor $\phi(\xi_{11}r/a)$ is given by (neglecting also in this case additive terms not depending on position):

\begin{equation}\label{phase_spher}
\tan\phi(\xi_{11}r/a)=\frac{r_{S}\hat{N}_{1}(\xi_{11}r/a)}{\hat{J}_{1}(\xi_{11}r/a)+r_{S}\hat{N}_{1}(\xi_{11}r/a)}=\frac{r_{S} \ y_{1}(\xi_{11}r/a)}{j_{1}(\xi_{11}r/a)+r_{S}y_{1}(\xi_{11}r/a)}
\end{equation}

We remind that in the case of spherical cavity we have $r_{S}=R_{S}/Z_{0}$.

\
Similarly to section II, also in this case the line equation yields for the current mode $I(r)$ the expression:

\begin{equation}\label{current_mode_spher0}
\frac{dV(r)}{dr}= -jZ_{0}k I(r)
\end{equation}

$k$ being the free-space wave number at the resonance, $k=2\pi/\lambda_{0}=\xi_{11}/a$ for the TE$_{110}$ mode. Inserting Eq. (\ref{mode_voltage_spher2}) into Eq. (\ref{current_mode_spher0}) we have:

\begin{equation}\label{current_mode_spher}
I(x)= j\frac{V_{0}}{Z_{0}}x \left[j_{0}(x)-\frac{j_{1}(x)}{x}\right] \ e^{j\left[\phi(x)+\psi(x)\right]}
\end{equation}

where for convenience we defined $x=\frac{\xi_{11}}{a}r$ (not to be confused with the Cartesian coordinate); the phase factor $\psi$ is given by:

\begin{equation}\label{psi}
\tan\psi(x)=\frac{\frac{d\phi}{dx}j_{1}(x)}{j_{0}(x)-\frac{1}{x}j_{1}(x)}
\end{equation}

The previous equations allow us to write the field components of the electromagnetic mode. In particular, defining $A_{H}$ as the value of $H_{\theta}$ in the cavity center and inserting Eqs. (\ref{mode_voltage_spher2}) and (\ref{current_mode_spher}) into Eq. (\ref{modes_te10_spher}) we have:

\begin{subequations}
\label{modes_te110}
\begin{eqnarray}
E_{\varphi}&=&\frac{3}{2} j A_{H} Z_{0} \sin\theta j_{1}(x)e^{j \phi(x)}\label{equationa}
\\
H_{\theta}&=& \frac{3}{2} A_{H} \sin\theta\left[j_{0}(x)-\frac{j_{1}(x)}{x}\right]  e^{j\left[\phi(x)+\psi(x)\right]}\label{equationb}
\\
H_{r}&=& -3 A_{H} \cos\theta \frac{j_{1}(x)}{x} e^{j\phi(x)}\label{equationc}
\end{eqnarray}
\end{subequations}

In the applications related to atomic frequency standards, the field component of interest is parallel to the quantization axis defined by a static magnetic field. The quantization axis is chosen along the $z$-axis so that:

\begin{equation}\label{h_z}
\begin{split}
H_{z}=&H_{r}\cos\theta-H_{\theta} \sin\theta\\
=&-A_{H}\left\{2\cos^{2}\theta \frac{j_{1}(x)}{x}+\sin^{2}\theta\left[j_{0}(x)-\frac{j_{1}(x)}{x}\right]e^{j\psi(x)}\right\}e^{j\phi(x)}
\end{split}
\end{equation}

It is interesting to notice that in the case of the spherical cavity both $\phi(x)$ and $\psi(x)$ play a role in the analysis of phase-shift, as it will be discussed in the following.

\
In Fig. 6 we report on the behavior of $\phi(z)$ for two values of $r_{S}$ for a spherical cavity resonating at $\nu_{0}=7.2$ GHz on the TE$_{110}$ with radius $a= 3$ cm.

\

\begin{figure}[!]
\begin{center}
\includegraphics[height=250pt]{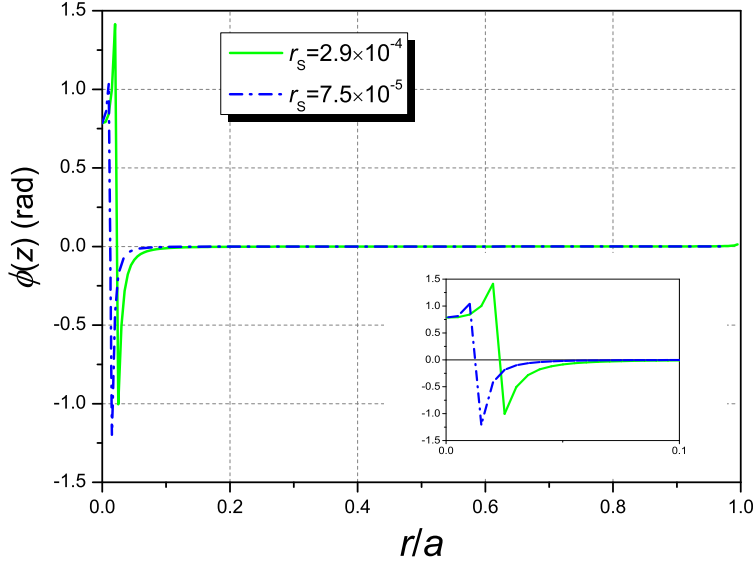}
\caption{(Color online) Plot of $\phi(z)$ versus $r/a$ for two values of $r_{s}$; dash-dot line: $r_{s} =7.5 \times 10^{-5}$ corresponding to a spherical cavity in Al; continuous line: $r_{s} =2.9 \times 10^{-4} $ corresponding to a spherical cavity in Ti. the inset shows the behavior of $\phi(z)$ around the singularity close to the cavity origin.}\label{tanphi_spher_cavity}
\end{center}
\end{figure}

Except for the regions around the center of the cavity and close to $r=a$, Eq. (\ref{phase_spher}) can be simplified as:

\begin{equation}\label{approx_phi}
\phi(x)=r_{S}\frac{y_{1}(x)}{j_{1}(x)}
\end{equation}

The phase factor $\psi(x)$ can be simplified as well; in particular, in the region $0<x<\xi_{11}$, Eq. (\ref{approx_phi}) can replace $\phi(x)$ in Eq. (\ref{psi}) so that we have:

\begin{equation}\label{approx_psi}
\tan\psi=\frac{r_{S}}{x j_{1}(x)}\frac{1}{\frac{d}{dx}\left[xj_{1}(x)\right]}
\end{equation}
and the corresponding behavior is reported in Fig. 7.

\
\begin{figure}[!]
\begin{center}
\includegraphics[height=250pt]{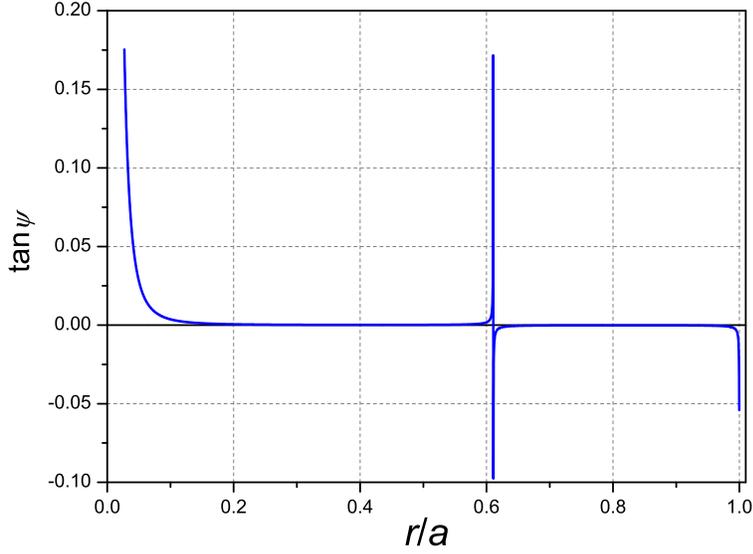}
\caption{(Color online) Plot of $\tan\psi$ versus $r/a$ for  $r_{s} =7.5 \times 10^{-5} $ (cavity in Al).}\label{tanpsi_spher_cavity}
\end{center}
\end{figure}
\
The singularity of $\tan\psi$ corresponds to the first zero of the function $\frac{d}{dx}\left[xj_{1}(x)\right]=x\left[j_{0}(x)-\frac{j_{1}(x)}{x}\right]$ that is $x=\xi_{11}^{TM}=2.7437$, $\xi_{11}^{TM}$ being an eigenvalue of the TM$_{110}$ mode, and corresponds to the sign inversion of $H_{\theta}$ for the TE$_{110}$, as we can notice from Eq. (\ref{modes_te110}).

\
The phase $\Phi_{z}$ of the field $H_{z}$ ($\arg H_{z}$), that is particularly important for frequency standards applications, can be
deduced from Eq. (\ref{h_z}) and in general will be a function of both $x$ and $\theta$: $\Phi_{z}=\Phi_{z}(x, \ \theta)$. In particular, on the $z-$axis ($\theta=0$) and on the equatorial plane $\theta=\pi/2$ its expression is given by:

\begin{equation}\label{phi_z}
\begin{split}
&\Phi_{z}(x, \ 0)= \phi(x)\\
&\Phi_{z}(x, \ \pi/2)=\phi(x)+\psi(x)
\end{split}
\end{equation}

Equation (\ref{h_z}) allows to write the complete expression for $\Phi_{z}(x, \theta)$; the argument of $H_{z}$ is given by:

\begin{equation}\label{soluzione}
\tan\Phi_{z}(x, \ \theta)= \frac{\Im H_{z}}{\Re H_{z}}=\frac{\cos^{2}\theta \frac{2j_{1}(x)}{x}\sin\phi(x)+\sin^{2}\theta \left[j_{0}(x)-\frac{j_{1}(x)}{x}\right]\sin(\phi(x)+\psi(x))}{\cos^{2}\theta \frac{2j_{1}(x)}{x}\cos\phi(x)+\sin^{2}\theta \left[j_{0}(x)-\frac{j_{1}(x)}{x}\right]\cos(\phi(x)+\psi(x))}
\end{equation}

\begin{figure}[!]
\begin{center}
\includegraphics[height=250pt]{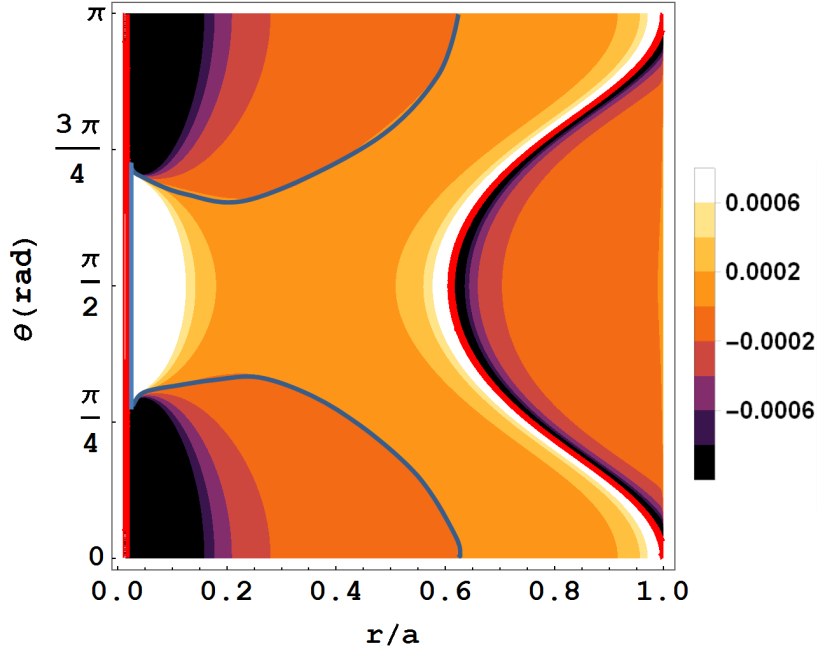}
\caption{(Color online) Contour plot of $\Phi_{z}$ versus $r/a$ and $\theta$ for  $r_{s} =7.5 \times 10^{-5} $ (cavity in Al).  The red line shows the points where the phase is singular, whereas the blue line shows the points where the phase is zero. The numbers reported in the colour legend on the right indicate the values of $\Phi_{z}$ in radians.}\label{tanphi_z_spher_cavity}
\end{center}
\end{figure}

Figure \ref{tanphi_z_spher_cavity} shows the contour plot of $\Phi_{z}(x, \theta)$. The red line represents the locus of points where the phase is singular, that is where it abruptly changes sign; the cavity origin is then a critical point, regardless the value of $\theta$. The figure shows that there is another bend of values where the phase is singular, even though they are normally not interested by clock operation.

\newpage
\section{Phase-shift in compact atomic clocks}

Compact atomic frequency standards based on either thermal or cold atoms have reached the best performances in pulsed operation and adopting the Ramsey technique to interrogate the atoms \cite{godone2015}. During the Ramsey interrogation phase, the atomic vapor can be considered as a two-level system coupled to an oscillating magnetic field sustained by a microwave cavity resonating at a frequency $\omega_{0}$ close to the atomic frequency $\omega_{hfs}$ (see Fig. 9(a)). For our purposes, the atomic frequency is defined by the ground-state hyperfine transition of Rb or Cs. As previously mentioned, the oscillating magnetic field component of interest is that parallel to the quantization magnetic field direction.

\
In Fig. 9(b), we show a Ramsey interaction scheme in the time domain, where $b$ is the (angular) microwave Rabi frequency defined as $b=\mu_{z} \mu_{0}H_{z}/\hbar$, $\mu_{z}$ being the magnetic dipole moment of the hyperfine transition and $\mu_{0}$ the permeability of free space. Throughout the paper we assume $t_{1}, \ t_{2}\ll T$, a condition well satisfied in experiments. The amplitudes of the two microwave pulses are assumed different: in this way, we take into account that the atoms can see different intensities of the oscillating magnetic field because of their motion inside the cavity during the Ramsey time.

\begin{figure}[!]
\begin{center}
\includegraphics[height=350pt]{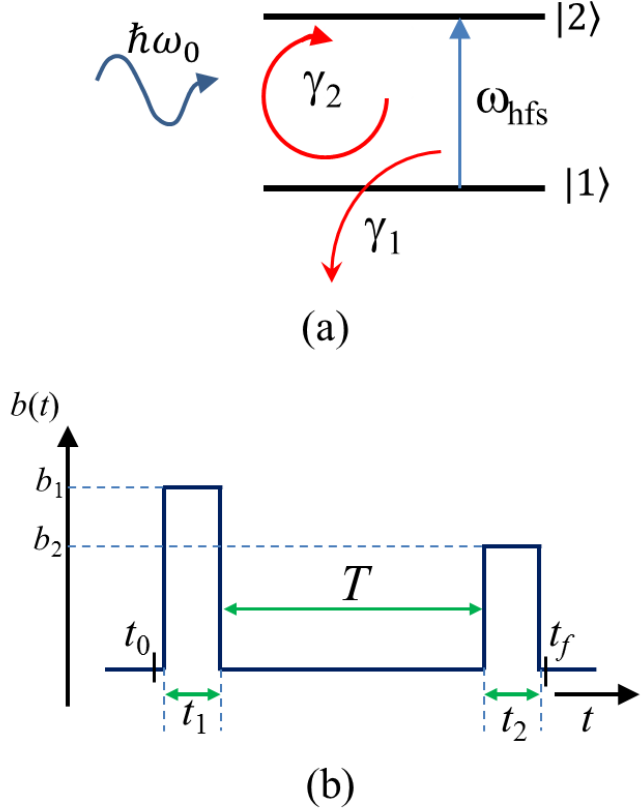}
\caption{(Color online) (a) Two-level system involved in clock operation; $\gamma_{1}$ and $\gamma_{2}$ are the longitudinal and transversal relaxation rates respectively; they are mainly related to the buffer gas interaction. (b) timing sequence of the Ramsey interaction scheme with two unequal amplitude microwave pulses.}\label{clock_ope}
\end{center}
\end{figure}
\
The population difference $\Delta$ between the two clock levels at the end of the Ramsey interaction ($t=t_{f}$) is the physical observable commonly used to detect the clock transition.

\
The central fringe of the Ramsey pattern as a function of the microwave detuning $\Omega_{\mu}\equiv\omega_{0}-\omega_{hfs}$ can be written with good approximation as:

\begin{equation}\label{delta_finale}
\Delta(\Omega_{\mu})=\Delta_{t_{0}}\left\{e^{-\gamma_{2}T}\sin\beta_{1}\sin\beta_{2}\cos\left[\Omega_{\mu}T+(\phi_{2}-\phi_{1})\right]+e^{-\gamma_{1}T}\cos\beta_{1}\cos\beta_{2}\right\}
\end{equation}

where $\Delta_{t_{0}}$ is the population difference at $t=t_{0}$ (beginning of the interaction), $\beta_{1}=b_{1}t_{1}$ ($\beta_{2}=b_{2}t_{2}$) is the microwave pulse area and $\phi_{1}$ ($\phi_{2}$) is the microwave phase experienced by the atoms during the first (second) Rabi pulse, $\gamma_{1}$ and $\gamma_{2}$ are the relaxation rates for population and coherence respectively when the atomic sample is placed in a cell with buffer gas. For cold-atoms $\gamma_{1}$ and $\gamma_{2}$ can be set to zero.

\
From previous equation, it is evident that a phase difference $\Delta\phi=\phi_{1}-\phi_{2}$ induces a frequency shift $\Delta \omega_{PS}$ of the clock transition whose fractional value is given by:

\begin{equation}\label{phase_shift}
\frac{\Delta \omega_{PS}}{\omega_{hfs}}=-\frac{\Delta\phi}{\omega_{hfs}T}=-\frac{\Delta\phi}{\pi Q_{a}}
\end{equation}

$Q_{a}=\omega_{hfs}T/\pi$ being the quality factor of the central Ramsey fringe.

\
In the following, we intend to estimate $\Delta \omega_{PS}$ for some experimental configurations; in particular, we are interested to use the relations we obtained in previous sections to evaluate $\Delta \omega_{PS}$ as results from the joint action of atomic motion and spatial variation of the microwave phase. The phase variations due to the phase noise of the microwave signal (Dick effect) are not considered in this work since extensively reported in the literature \cite{santarelli1998, audoin1998, dick1987}.

\
Equations (\ref{delta_finale}) and (\ref{phase_shift}) refer to a single atom; for a sample of atoms with a finite size inside the cavity, the total behavior will be a spatial average of single contributions. Such an average is complicated by the fact that the atoms move during the Ramsey interaction. However, when the atomic motion is predictable, for example when due to gravity or to thermal expansion, it is possible to find an expression for the average phase-difference $\langle\Delta\phi\rangle$ experienced by the cloud and for the corresponding phase-shift (see appendix).

\subsection{Cold atoms}

For atomic clocks based on isotropic cooling, the atomic vapor is a molasses with thermodynamic temperature in the range $5-100 \ \mu$K, corresponding to $|\vec{v}|< 10 \ \textrm{cm}/\textrm{s}$.

\
We evaluate the phase-shift for two situations of interest for cold-atom clocks applications: clock in a 0-g environment (space clock) and clock on the ground.
The change of position of the atoms inside the cavity during the Ramsey period $T$ is then either related to the molasses expansion (for a space clock) or mainly to the free fall of the molasses due to gravity (for a ground clock).

\subsubsection{Cylindrical cavity}
We initially consider the case of a clock in 0-g condition. A spherical molasses of $^{87}$Rb atoms is generated in the center of a cylindrical cavity operating on the TE$_{011}$ mode. This assumption is supported by the experimental evidence that with appropriate configuration of injection
lasers the cold atoms can be collected in the middle region of the cavity in a Gaussian-like shape \cite{meng2013}.

\
After the first microwave pulse, the cloud is supposed to expand due to the residual thermal velocity for a Ramsey time $T$. Specifically, if at the beginning of the interaction the molasses conventional radius is $r_{m1}$ (see appendix), at the end of the Ramsey period it will be $r_{m2}=r_{m1}+\bar{v} T$, where $\bar{v}$ is the thermodynamic speed of the cold atomic vapor. Then, the atoms are interrogated again and detected. In particular, as explained in the appendix, we suppose to perform the detection through a laser beam with a propagation direction along the $z$ axis; therefore, only the atoms interacting with the detection laser contribute to the signal. We assume a detection laser with a section of 10 mm in diameter.

\begin{figure}[!]
\begin{center}
\includegraphics[height=200pt]{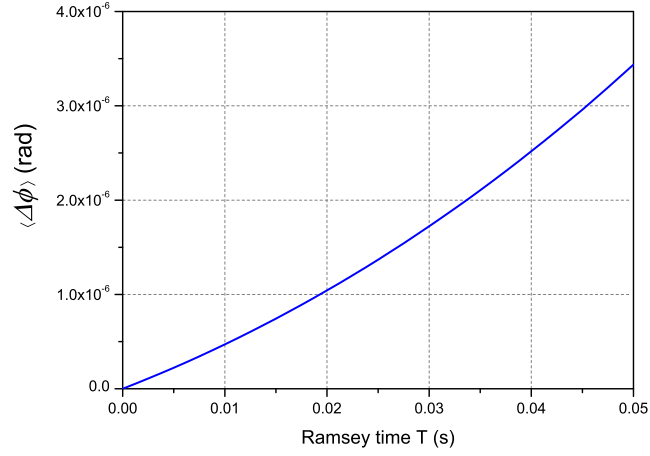}
\caption{$\langle\Delta\phi\rangle$ versus $T$ for an expanding molasses of Rb atoms in a cylindrical cavity made in Al. Molasses temperature: 30 $\mu$K.}\label{phase_cylindrical_cavity_expansion}
\end{center}
\end{figure}

\
Figure \ref{phase_cylindrical_cavity_expansion} shows the average phase difference $\langle\Delta\phi\rangle$ seen by the atom-cloud and calculated according to the procedure reported in the appendix. Figure \ref{cylindrical_cavity_expansion2} reports the induced frequency shift; both the figures refer to a molasses temperature of 30 $\mu$K (corresponding to $\bar{v}\approx 7.5 \ \textrm{cm}/\textrm{s}$). Since in clock applications we are interested to long Ramsey times, the phase-shift is calculated for $T>10$ ms.

\begin{figure}[!]
\begin{center}
\includegraphics[height=200pt]{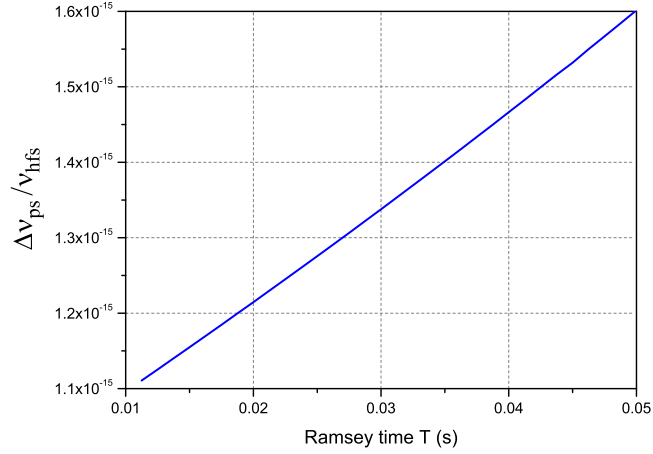}
\caption{Phase-shift for the same situation of previous figure.}\label{cylindrical_cavity_expansion2}
\end{center}
\end{figure}

The interpretation of these two figures is clear. According to Fig. 4, the behavior of the phase around the cavity center where the atoms are located  is regular, there are no singularities and since the cloud expands isotropically around the origin the corresponding shift is small, at the level of $10^{-15}$.

\
As a second example, we consider a ground-based clock. Also in this configuration the molasses is supposed to be placed in the cavity center at the beginning of the interaction. The quantization axis and the gravity vector are parallel and oriented along the $z$-axis. The molasses radius is $r_{m}$ and the change of position during the Ramsey time is mainly due to gravity: all the cloud moves along the $z$-axis of a quantity $\Delta z=\frac{1}{2}g T^{2}$, $g$ being the gravity acceleration (in this case we disregard the expansion in the evaluation of the phase-shift).

\begin{figure}[!]
\begin{center}
\includegraphics[height=200pt]{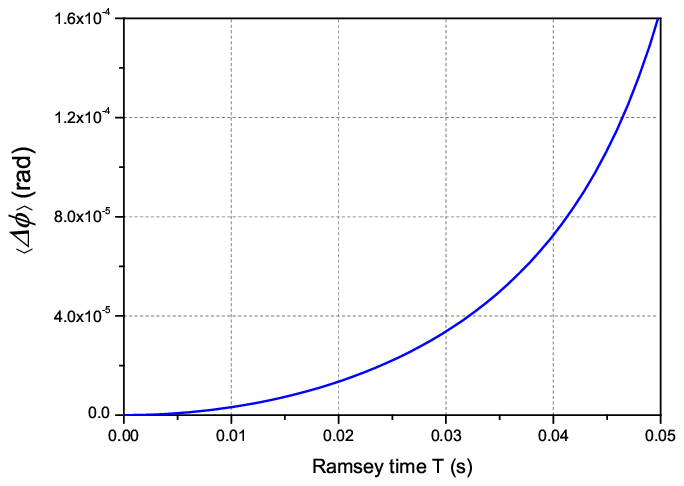}
\caption{$\langle\Delta\phi\rangle$ versus $T$ for a falling molasses of Rb atoms in a cylindrical cavity made in Al.}\label{phase_cylindrical_cavity_fall}
\end{center}
\end{figure}

\begin{figure}[!]
\begin{center}
\includegraphics[height=200pt]{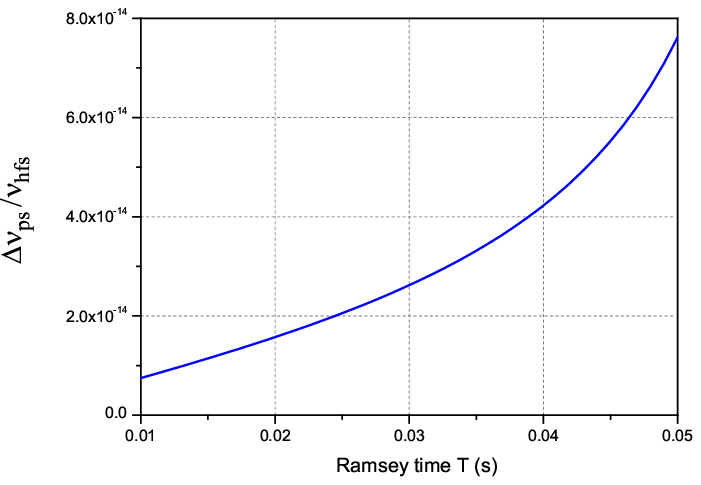}
\caption{Phase-shift corresponding to $\langle\Delta\phi\rangle$ of the previous figure.}\label{cylindrical_cavity_fall}
\end{center}
\end{figure}

Similarly to the previous case, figures \ref{phase_cylindrical_cavity_fall} and \ref{cylindrical_cavity_fall} show $\langle\Delta\phi\rangle$ and the associated phase-shift.

\
In the configuration of ground clock, the frequency shift due to the phase variations can reach units of $10^{-14}$. However, we notice that, within the approximations done so far, the phase-shift can be made negligible when the direction of the gravity acceleration is orthogonal to $\hat{\textbf{z}}$, since the phase variations depend on $z$ only.

\
A comparison between the two situations shows that the phase-shift is less critical for the expansion, though not completely negligible, than for the fall in the gravitational field.

\subsubsection{Spherical cavity}

In this subsection, we consider the case of a spherical molasses inside a spherical cavity operating in the TE$_{110}$, close to the experimental setup described in \cite{esnault2010, wang2012, esnault2007}. As reported in \cite{wang2012, esnault2007}, the spatial distribution of atoms exhibits two lobes; specifically, only few atoms are located in the cavity center and most of them are approximately in the middle between the center and inner surface of the sphere. However, during the cooling phase it is possible to implement a ”Sisyphus like” technique by ramping down to zero the laser cooling intensity and by increasing its detuning, as explained in \cite{esnault2007}. In this case, temperatures of the order of a few $\mu$K are reached and the atoms are grouped in the cavity center.

\
Under this assumption, the evaluation of the phase-shift can be done following the approach used for the cylindrical cavity. The main difference is that for the spherical cavity the phase of $H_{z}$ is singular in the region close to the origin, exactly where the molasses is formed (see Figs. 6,7, 8).
To clarify what happens in this case, it is useful to study the behavior of a single atom.

\
We consider a space clock and one atom placed near the origin on the equatorial plane and belonging to an expanding molasses. Due to expansion, we suppose the atom continues to move radially on the equatorial plane so that it can eventually come across a second phase singularity for $ r \approx 0.6 a$ (see Fig. 8). This means that the atom might experience a phase $\approx +\pi/2 $ during the first microwave pulse and  $\approx -\pi/2 $ during the second one, so that according to Eq. (\ref{phase_shift}) $\Delta\phi \approx \pi$ and the contribution to the phase-shift given by this atom might be as large as $2 \times 10^{-10}$ .

\
The situation is very different for one atom of the cloud still moving along the equatorial plane but far from the singularities; in this case we can refer to Eq. (\ref{phi_z}) and use the simplified expressions Eqs. (\ref{approx_phi}) and (\ref{approx_psi}) for $\phi(x)$ and $\psi(x)$, respectively. Assuming that during the interaction the atom moves from $r=0.5$ cm to $r=1$ cm (cavity radius $a=3.17$ cm), the relative frequency shift induced by the phase variation is $-3.8\times 10^{-14}$.

\
For a more realistic sample of cold atoms initially contained in a molasses of radius $r_{m}=0.5$ cm, we have to average over the entire sample, as explained in the appendix. It is interesting to compare different molasses temperatures and different cavity materials.

\begin{figure}[!]
\begin{center}
\includegraphics[height=350pt]{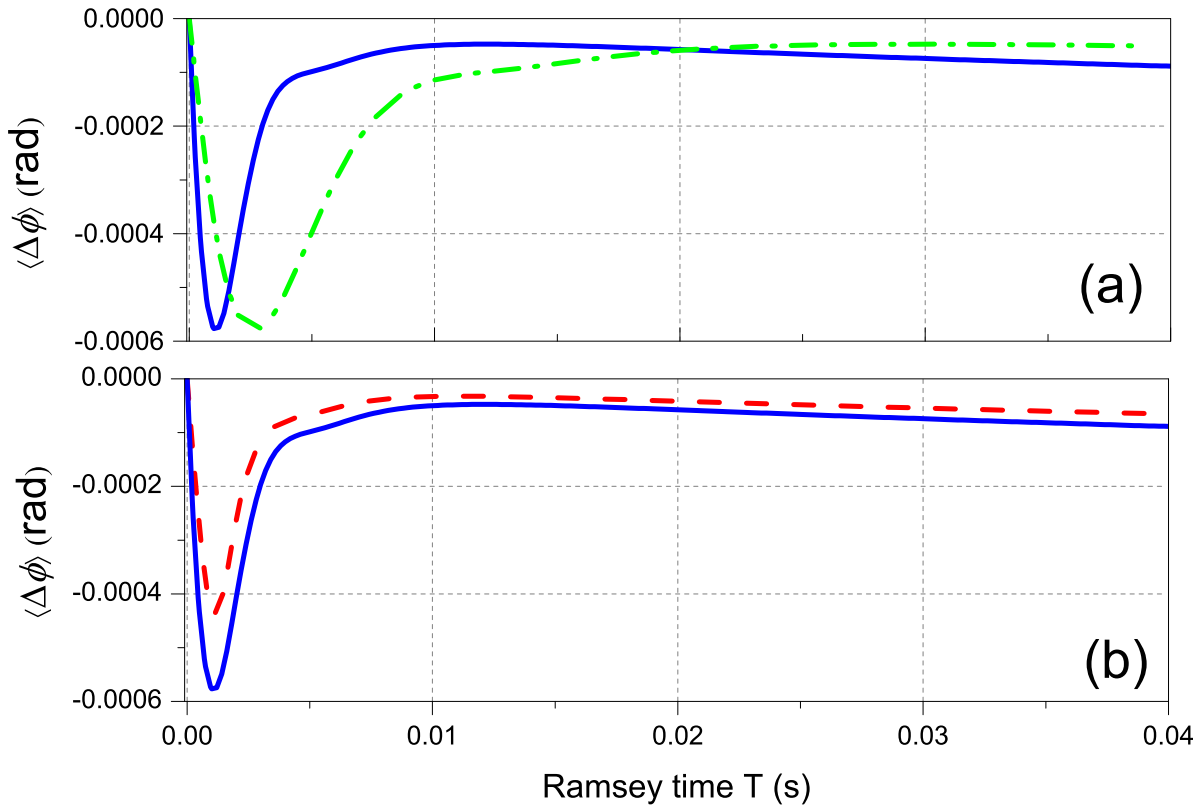}
\caption{(Color online) $\langle\Delta\phi\rangle$ versus the Ramsey time $T$; (a) cavity in Al for two different temperatures of the molasses; Continuous line: 30 $\mu$K; dash-dot line: 5 $\mu$K; (b) temperature 30 $\mu$K for two different cavity materials; continuous line: Al; dash line Cu}\label{delta_phi_exp_sphere}
\end{center}
\end{figure}

\begin{figure}[!]
\begin{center}
\includegraphics[height=200pt]{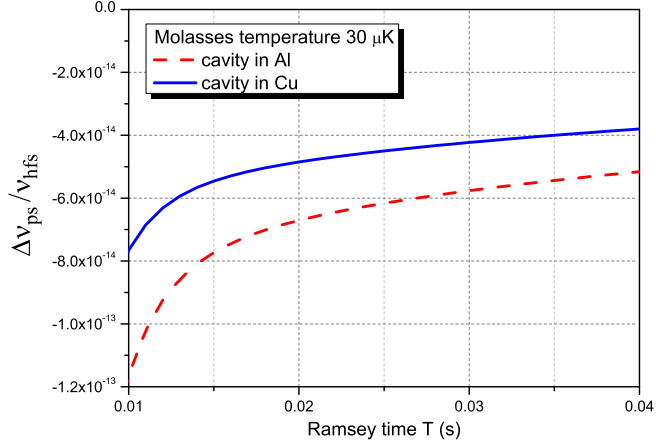}
\caption{(Color online) Phase-shift for the values of the parameters corresponding to Fig. \ref{delta_phi_exp_sphere}b}\label{shift_cu_al}
\end{center}
\end{figure}

\
Results are shown in Figs. \ref{delta_phi_exp_sphere} and \ref{shift_cu_al} for a $^{87}$Rb molasses.

Figure \ref{delta_phi_exp_sphere} shows $\langle\Delta\phi\rangle$  versus the Ramsey time $T$; in (a) we consider a cavity in Al for two different temperatures of the molasses; in (b) we report $\langle\Delta\phi\rangle$ for two materials (Al and Cu) with different losses. The peak corresponds to the atoms that, starting from the origin, reach the singularity after a few milliseconds. Reasonably, the colder is the atomic cloud, the later the atoms reach the singular point. On the other hand, the peak height depends on losses, not on molasses temperature, so it is larger in Al ($r_{S}=7.5 \times 10^{-5}$) than in Cu ($r_{S}=5.6 \times 10^{-5}$).

\
Figure \ref{shift_cu_al} shows the phase-shift corresponding to Fig. \ref{delta_phi_exp_sphere}(b) for $T> 10$ ms. In that range, the average phase-difference is nearly constant and consequently the phase-shift simply scales as $1/T$. To give an order of magnitude, for $T= 40$ ms and a cavity in copper the relative fractional phase shift is $-3.8 \times 10^{-14}$.

We point out that to evaluate the phase shift we applied Eq. (\ref{soluzione}) using for $\phi(x)$ and $\psi(x)$ the expressions not simplified (Eqs. (\ref{phase_spher}) and (\ref{psi})).

\
We now consider a ground clock where the atoms undergo the Ramsey interaction in free fall, under the action of gravity. Also in this case it is useful to examine the contribution to the phase-shift of a single atom of the cloud; specifically, we suppose the atom falls along the $z-$axis. In this case the atom only feels the singularity near the origin, no other abrupt phase leaps are present along the atom's trajectory. For a Ramsey time $T= 45$ ms, corresponding to a gravity fall of $\Delta z=1.$ cm, we find a phase-shift of $4\times 10^{-10}$.

\begin{figure}[!]
\begin{center}
\includegraphics[height=200pt]{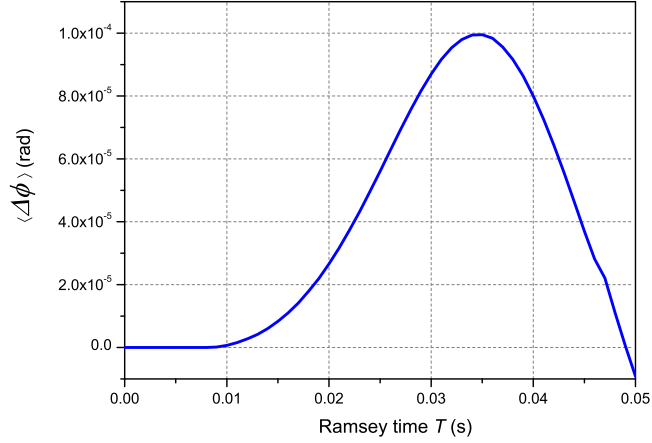}
\caption{$\langle\Delta\phi\rangle$ versus Ramsey time $T$ for a Rb molasses falling under the action of gravity in a cavity made in Al.}\label{phase_sphere_fall}
\end{center}
\end{figure}

\begin{figure}[!]
\begin{center}
\includegraphics[height=200pt]{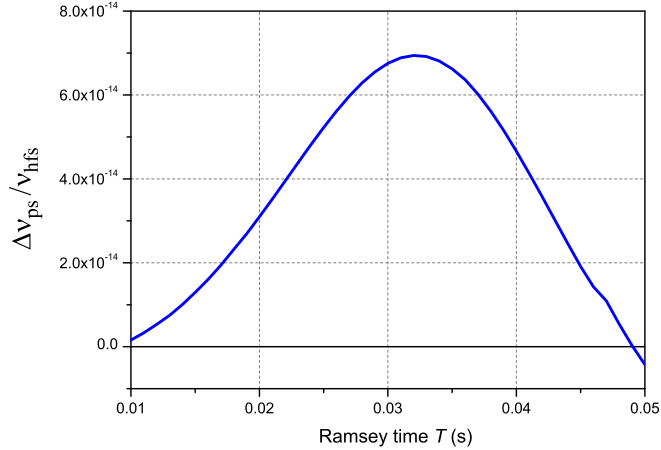}
\caption{Phase-shift for the same configuration of the previous figure.}\label{sphere_fall0}
\end{center}
\end{figure}

This value can be averaged down taking into account the entire molasses that, as in the previous case, we suppose to have a radius of $r_{m}=0.5$ cm and placed initially at the origin of the cavity. In Fig. \ref{phase_sphere_fall} we report the phase-variations averaged over the spatial distribution in the molasses and in Fig. \ref{sphere_fall0} the corresponding shift.

\
We notice that for small Ramsey times, $\langle\Delta\phi\rangle$ remains almost negligible because the molasses radius is large compared to the singularity region, therefore the effects of the rapid phase leap are averaged out. Increasing the Ramsey time, also the averaged phase increases and eventually reaches a peak, corresponding to the time spent by the cloud to travel a length approximatively equal to its radius. Indeed, for this Ramsey time, there are no more atoms to compensate for the phase jump experienced by the atom cloud during the first microwave pulse.

\
The phase-shift represented in Fig. \ref{sphere_fall0} follows the behavior of $\langle\Delta\phi\rangle$.  Notably, we point out that a Ramsey time may exist which makes the phase-shift negligible.

\subsection{Thermal atoms}

In this subsection, we provide an estimate of the phase-shift in vapor cell frequency standards operating with hot atoms diluted in a buffer gas.

\
The average time $\tau_{D}$ between two consecutive collisions of an atom with buffer gas atoms/molecules is given by \cite{Vanier}:

\begin{equation}
\tau_{D}=\frac{M D_{0}}{k_{B}T_{0}}\frac{P_{0}}{P}
\end{equation}

where $M$ is the atom mass, $k_{B}$ is the Boltzmann's constant, $T_{0}$ the buffer gas temperature, $D_{0}$ the diffusion coefficient, $P_{0}$ the reference pressure ($P_{0}=760$ Torr) and $P$ the buffer gas pressure in the cell. Assuming for the atom an average free path between two consecutive collisions of $l_{m}=\bar{v}\tau_{D}$  ($\bar{v}$ is the atom's average thermal speed) and a random walk motion (Wiener-Levy process), the typical atom's change of position $\Delta \vec{s}$ between two Rabi pulses can be written as:

\begin{equation}
|\Delta \vec{s}|=\sqrt{\frac{T}{\tau_{D}}}l_{m}=\sqrt{\frac{8}{\pi}\frac{D_{0} P_{0}T}{P}}
\end{equation}

\
For the POP $^{87}$Rb frequency standard described in \cite{micalizio2012} we have $T_{0} = 60 \ ^{\circ}$ C, $\bar{v}= 270 \ \textrm{m}/\textrm{s}$, $P=25$ Torr, $D_{0}=0.25 \ \textrm{cm}^{2}/\textrm{s}$ and $T= 2.6$ ms, so that we find $\tau_{d}= 26$ ns, $l_{m}= 7 \ \mu$m and  $\Delta \vec{s}=2.2$ mm. In the experimental setup it was used a cylindrical cavity (resonating on TE$_{011}$ mode) in Molybdenum ($R_{S}=40$ m$\Omega$) with short-circuits (end-caps) in Aluminium ($R_{S}= 29$ m$\Omega$) and with size $2a= 52$ mm and $d=49$ mm, for which we have $\alpha_{i}=0$, $\alpha_{w}=8.6 \times10^{-3} \ \textrm{m}^{-1}$ (attenuation coefficient for the waveguide in Mo), and $r_{S}=3.1 \times10^{-6}$ (normalized surface resistance of the end-caps).

\
Using Eqs. (\ref{phase_2}) and (\ref{phase_shift}) we can estimate $\Delta \omega_{PS}$ for one atom diffusing in the buffer gas and traveling a path  $\Delta z=\Delta \vec{s}\cdot \hat{z}$; in the most conservative case $\Delta z=|\Delta\vec{s}|$, and for an atom initially in $z=0$ (center of the cavity)
) Eq. (\ref{phase_shift}) gives a relative phase-shift of $\Delta\nu/\nu\approx -2\times10^{-13}$. Similarly, for an atom initially in $z=-10$ mm during the first microwave pulse we have $\Delta\nu/\nu\approx -5\times10^{-13}$.

\
These frequency shifts are large for an atomic frequency standard aiming to achieve a frequency stability in the range of $10^{-15}$; however, the average over the spatial distribution of the atomic vapor makes negligible the phase-shift: for each atom moving along the positive $z$-axis there is one atom moving in the opposite direction, so that the overall effect is null. The phase-shift in a cell system is then related to possible non-uniformity of the spatial distribution of the atomic medium. These non-uniformities can be due to surface effects on the inner walls of the cell, to non uniform deposition of Rb (or Cs) in the walls, to temperature gradients in the cell, or more generally to a non-equilibrium behavior of the Rb vapor diffusing in the buffer gas. The study of these effects is beyond the goal of the present paper. We report here a qualitative example to give a rough estimate of the phase-shift.

\
Very simply, we suppose that the atoms moving in one direction travel a length $\Delta\zeta_{1}=2.2$ mm, as previously calculated, whereas those moving in the opposite direction a length $\Delta\zeta_{2}= 2.17$ mm. Following the calculation reported in the appendix, this difference in the two paths (which is of the order of 1\%) produces a phase-shift of $-2\times 10^{-15}$. This result suggests that spatial non-uniformities of a few percent may play a role in limiting the frequency stability of high-performing cell clocks.

\section{Conclusions}

In this work we have studied the spatial phase-shift in compact atomic clocks due to losses in the resonant cavity.

\
We have written the quasi-stationary fields that are of particular interest in clock applications: the TE$_{011}$ mode of a cylindrical cavity and the TE$_{110}$ of a spherical cavity. In the expressions of the fields, we have specified the phase factor associated to the steady state wave; this phase factor accounts for the (low) losses in the cavity.

\
We have provided an order of magnitude of the frequency shift induced by the spatial variation of the phase on the clock transition, both in cold- and thermal-atom clocks.

\
Since cold-atoms frequency standards are intended to also provide the accuracy specification, a more precise evaluation of the phase shift could require to take into account the nature of the cavity feeds, the real cavity geometry, the holes distribution and the spatial variation of $H_{z}$. Nevertheless, the theory here reported makes clear some fundamental aspects related to the cavity geometry. Specifically, the cylindrical cavity shows a phase singularity corresponding to the short-circuit end-caps, where usually the atoms are not present and where the intensity of the field is negligible. Instead, the spherical cavity exhibits a singularity close to the cavity center $r=0$, where the atoms are inevitably present and in other cavity regions where the field intensity is significant. In addition, cold atoms behave differently depending if they are placed in a cylindrical or in a spherical cavity. For the cylindrical cavity, it is possible to individuate a (unique) propagation direction (the $z$-axis) so that the phase shift can be significantly reduced by turning the $z$-axis orthogonally to the direction of the gravity acceleration. This is not possible for the spherical cavity, since the propagation direction of the traveling wave is not univocally defined.

\
We still point out that the phase-shift can also play a role in limiting the frequency stability of cold-atom clocks, being the phase-shift dependent on the position of the molasses in the cavity and on its shape and size.

\
Therefore, we argue that in \cite{pengliu} and in \cite{esnault2007, esnault2010}, the phase-shift effect may contribute to limit the long-term stability of the clocks.


\
Vapor-cell frequency standards based on thermal atoms are intrinsically not accurate and a precise evaluation of the phase-shift would be in principle not needed. However, similarly to the case of cold-atom clocks, the phase-shift can affect the frequency stability: the requirement of a stability in the $10^{-15}$ region may impose severe constraints on the non uniformity of the atomic vapor and on the presence of thermal gradients inside the cell.

\

\section{Appendix: Phase-shift for a finite-size atomic cloud}

In this appendix we present the method to extend the phase-shift calculation to a finite-size atomic sample.

\
We initially consider one atom of the sample that during the first microwave pulse is identified by the position vector $\mathbf{r}$ in a certain coordinate system. During the Ramsey time $T$ the atom moves so that it will be in $\mathbf{r'=r+r_{0}}$ when the atomic sample undergoes the second interrogation. The Ramsey signal produced by the atom at the end of the second pulse can then be written as a function of the microwave detuning $\Omega_{\mu}$ as:

\begin{equation}\label{delta_atom}
\Delta^{atom}(\Omega_{\mu})\propto\sin\left[\beta(\mathbf{r})\right]\sin\left[\beta(\mathbf{r'})\right]\cos\left[\Omega_{\mu}T+\phi(\mathbf{r'})-\phi(\mathbf{r})\right]
\end{equation}

In general, the extension of the previous equation to a sample of atoms is not easy due to the inability to model the motion of each atom of the sample. However, for cold atoms the motion becomes particularly simple since the atoms move coherently either under thermal expansion or by gravity fall. It is then easy to find a relation linking the coordinates $\mathbf{r'}$ that one atom takes during the second microwave pulse to the position $\mathbf{r}$ that the atom assumes during the first pulse and, most importantly, this relation applies to all the atoms of the cloud. If we then suppose that for a specific situation we have found $\mathbf{r'}=\mathbf{r'(r)}$, the Ramsey signal produced by the entire cloud can be written as:

\begin{equation}\label{signal_cloud}
\begin{split}
\Delta^{cloud} (\Omega_{\mu})& \propto \int d^{3}r \exp\left(-\frac{r^2}{r_{0}^{2}}\right) \sin\left[\beta(\mathbf{r})\right]\sin\left[\beta(\mathbf{r'(r)})\right] \times \\
&\times\cos\left[\Omega_{\mu}T+\phi(\mathbf{r'(r)})-\phi(\mathbf{r})\right]
\end{split}
\end{equation}

where we take into account that the atoms belong to a 3-D Gaussian spherically-symmetric cloud, $r_{0}$ being the conventional $e^{-1}$ radius of the cloud at the time $t_{0}$. Formally, the integral over the radial dimension extends to infinite.  However, to model an actual experiment, the integral is done over a finite size region; in particular, we suppose to perform the detection by means of a laser traveling along the $z$ axis of the cavity. Therefore, only the atoms that see the detection laser contribute to the clock signal.

\
Expanding in terms of $\Omega_{\mu}$ and looking for the maximum of the signal, Eq. (\ref{signal_cloud}) yields the following expression for the average phase difference experienced by the atoms of the cloud:

\begin{equation}\label{delta_phi_cloud}
\langle\Delta\phi\rangle=\arctan\left[\frac{\int d^{3}r \exp \left(-\frac{r^2}{r_{0}^{2}}\right)f(\mathbf{r})\sin\Phi(\mathbf{r})}{\int d^{3}r \exp\left(-\frac{r^2}{r_{0}^{2}}\right) f(\mathbf{r})\cos\Phi(\mathbf{r})}\right]
\end{equation}

where we defined $f(\mathbf{r})=\sin\left[\beta(\mathbf{r})\right]\sin\left[\beta(\mathbf{r'(r)})\right]$ and $\Phi(\mathbf{r})=\phi(\mathbf{r'(r)})-\phi(\mathbf{r})$. For the respective phase-shift we have then:

\begin{equation}\label{delta_shift_cloud}
\frac{\Delta\omega_{PS}}{\omega_{hfs}}=-\frac{\langle\Delta\phi\rangle}{T \omega_{hfs}}
\end{equation}

The previous equation point out that for an extended atomic cloud the clock signal is a non-trivial function of the field intensity experienced by each atom in a given position.

\
As an example of application of the previous analysis, we consider a molasses of cold-atoms in a spherical cavity and we evaluate $\langle\Delta\phi\rangle$ in the case of thermal expansion. We assume that during the first microwave pulse, one atom of the cloud is identified in spherical coordinates (see Fig. 9) by its radial distance $r$ from the origin, by the co-latitude $\theta$ ($0 \leq \theta \leq \pi$) and by the anomaly $\varphi$ ($0 \leq \varphi<2\pi$, however, for symmetry reasons, $\varphi$ does not play any role and will be disregarded in the following).

\begin{figure}[!]
\begin{center}
\includegraphics[height=350pt]{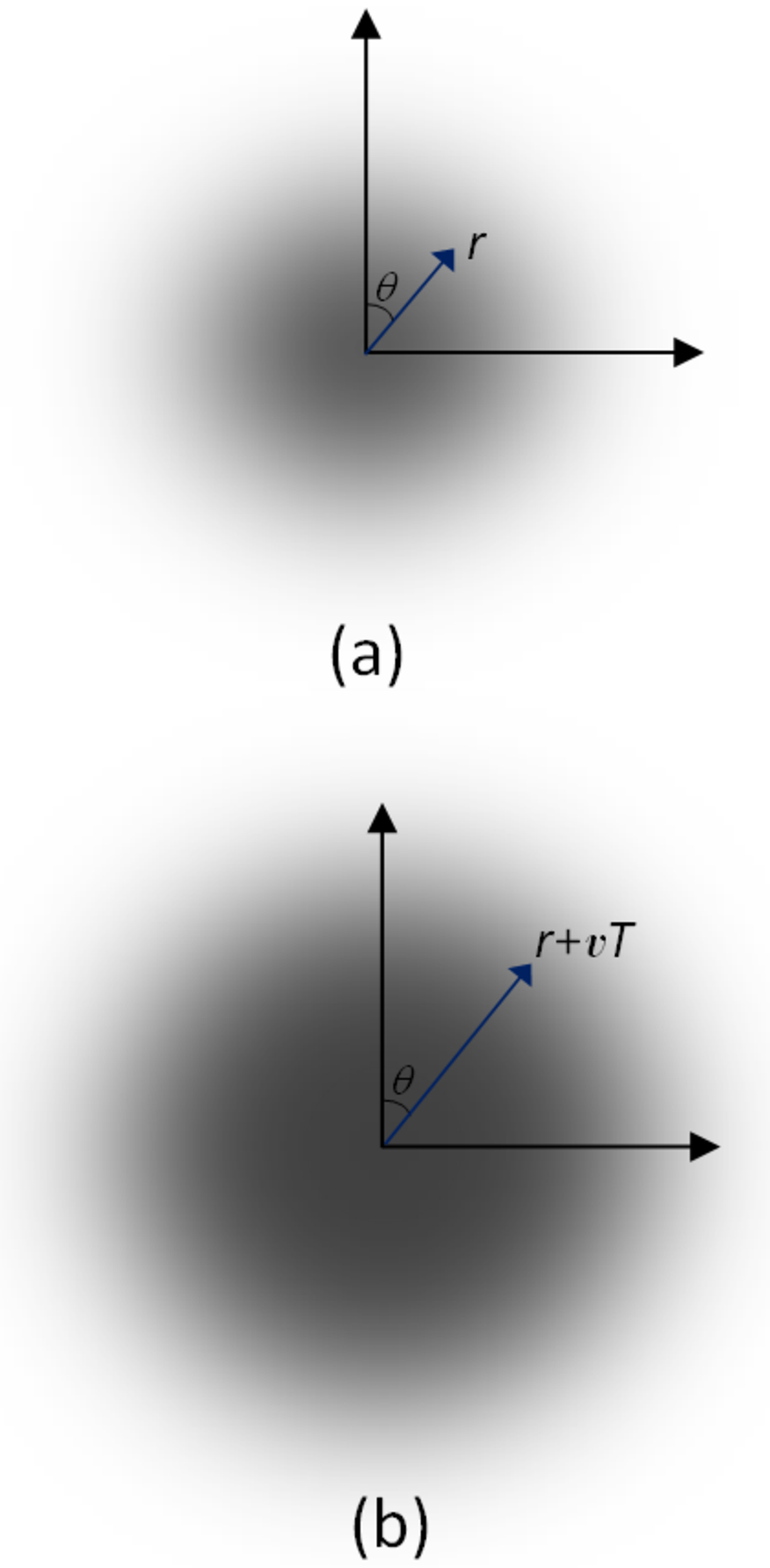}
\caption{$x-z$ section of an expanding cloud of cold atoms in 0-g condition; (a) the cloud experiences the first microwave pulse, a generic atom of the cloud is identified by $r$ and $\theta$; (b) after a time $T$ the cloud expands and experiences a second microwave pulses. The atom moves along the radial direction to the position $r+v T$.}\label{sphere_expansion}
\end{center}
\end{figure}

We suppose that due to expansion the atom moves radially at a speed $v$ for a time $T$ so that during the second microwave pulse its position is identified by $r+v T$ and by the same angular variables. The relation $\mathbf{r'}=\mathbf{r'(r)}$ is then simply given by:

\begin{equation}\label{trasf_coordinate1}
\begin{split}
r'&= r+vT\\
\theta' &=\theta
\end{split}
\end{equation}

Rigorously, we should take into account that the atoms' velocities are spread. However, in order to simplify the numerical integration we assume that all the atoms move at the most probable speed, as results from the Maxwell-Boltzmann distribution: $v=\sqrt{\frac{2 k_{B} \Theta}{m}}$, where $m$ is the mass of a single atom, $k_{B}$ the Boltzmann constant and $\Theta$ the temperature of the atomic cloud. A numerical integration of Eq. (\ref{delta_phi_cloud}) with Eq. (\ref{trasf_coordinate1}) gives $\langle\Delta\phi\rangle$ for an expanding cloud.

\
For a sample of cold atoms falling under the action of gravity, we consider Fig. 18 where the atoms are supposed to fall for a time $T$. Similarly to the previous case, we write the relation between the coordinates that a generic atom of the cloud takes during the first ($r, \ \theta$) and the second pulse ($r', \ \theta'$).

\begin{figure}[!]
\begin{center}
\includegraphics[height=350pt]{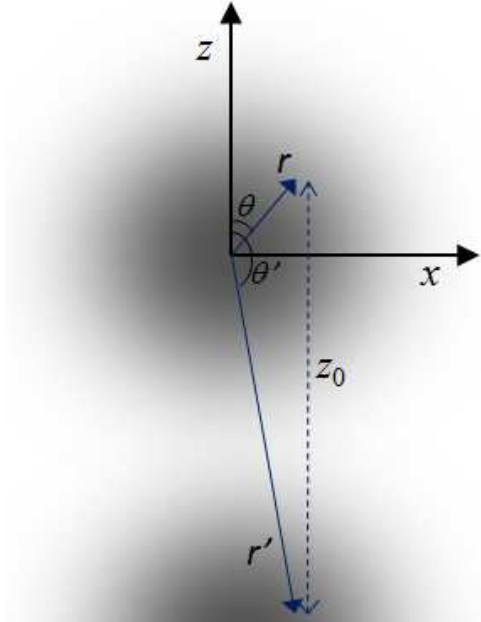}
\caption{$x-z$ section of a cold atoms cloud falling because of gravity. During the first microwave pulse the cloud is centered in the origin; after a time $T$ the cloud moved vertically.}\label{sphere_fall1}
\end{center}
\end{figure}

We have:

\begin{equation}\label{trasf2}
\begin{split}
r'&=\sqrt{r^2-2rz_{0}\cos\theta+z_{0}^2}\\
\theta'&=\pi-\arcsin\left[\frac{r\sin\theta}{\sqrt{r^2-2rz_{0} \cos\theta+z_{0}^2}}\right]
\end{split}
\end{equation}

where $z_{0}=\frac{1}{2}g T^2$ ($g$ being the gravity acceleration). The evaluation of $\langle\Delta\phi\rangle$ is done again through numerical integration of Eq. (\ref{delta_phi_cloud}). In the case of gravity fall we disregarded the effects related to thermal expansion.

\
The evaluation of the phase-shift for a cold sample of atoms in a cylindrical cavity in the cases of thermal expansion or gravity fall follows from analogous considerations.

\
For thermal atoms in a cylindrical cavity, the evaluation of the phase-shift according to the procedure here reported seems not applicable since the atomic motion is in this case totally random and in principle it is not possible to find any general relation $\mathbf{r'}=\mathbf{r'}(\mathbf{r})$ between the positions assumed by the atoms during the two microwave pulses. However, to provide an order of magnitude for the phase-shift also in this case, we can make some approximations. First, as explained in the text, we are interested to the phase-variation along the $z$-axis, so we suppose that all the atoms move either along the positive or negative direction of $z$-axis. Owing to the randomness of the atomic motion, half of the cell volume ($V/2$ in Eq. (\ref{delta_phi_cloud2})) is occupied by atoms traveling in one direction and the other half by those moving in the opposite direction.

\
Under this approximation, the phase variations averaged over the cell volume becomes:

\begin{equation}\label{delta_phi_cloud2}
\langle\Delta\phi\rangle=\arctan\left[\frac{\int_{V/2} d^{3}r f_{1}(\mathbf{r})\sin\Phi_{1}(z)+\int_{V/2}d^{3}r f_{2}(\mathbf{r})\sin\Phi_{2}(z)}{\int_{V/2} d^{3}r  f_{1}(\mathbf{r})\cos\Phi_{1}(z)+\int_{V/2} d^{3}r  f_{2}(\mathbf{r})\cos\Phi_{2}(z)}\right]
\end{equation}

where we defined:

\begin{equation}
f_{1}(\mathbf{r})=\sin\left[\beta_{1}J_{0}\left(\frac{x_{01}' r}{a}\right)\cos\left(\frac{\pi}{d}z\right)\right]\sin\left[\beta_{1}J_{0}\left(\frac{x_{01}' r}{a}\right)\cos\left(\frac{\pi}{d}(z+\Delta\zeta_{1})\right)\right]
\end{equation}

and

\begin{equation}
\Phi_{1}(z)=\phi(z+\Delta\zeta_{1})-\phi(z)
\end{equation}

 $\Delta\zeta_{1}$ being the path traveled by the atoms during the Ramsey time $T$ along the positive direction of $z$-axis and $\phi(z)$ is given by Eq. (\ref{phase_2}). Analogous definitions hold for $f_{2}(\mathbf{r})$, $\Phi_{2}(z)$, and $\Delta\zeta_{2}$.


\section*{Acknowledgements}

The authors would like to thank Filippo Levi for useful discussions. This work has been funded by the EMRP programme (IND55 MClocks). The EMRP is jointly funded by the EMRP participating countries within EURAMET and the European Union.

\end{document}